\documentclass[12pt]{article}
\usepackage{graphics}
\begin{document}
\title{JLC Overview\footnote{
Talk presented at APPI2002 workshop, 13-16 February, 2002, APPI,
Iwate, Japan}}
\author{Akiya Miyamoto\footnote{E-mail:akiya.miyamoto@kek.jp}\\
High Energy Accelerator Research Organization(KEK)\\
1-1 Oho, Tsukuba, Ibaraki 300-0805, Japan}
\date{}
\maketitle

\begin{abstract}
JLC is an $e^+e^-$ linear collider designed for experiments at $\sqrt{s}=500$ GeV 
with a luminosity of up to about $2.5\times 10^{34}/cm^2/s$. 
In this talk, after describing the parameters of JLC accelerator and detector, 
the feasibilities of JLC to study Higgs, Top, and SUSY physics are presented 
based on the ACFA report.
\end{abstract}

\section{Introduction}
JLC is a linear collider for $e^+e^-$ collision at the energy frontier.
The designed initial center-of-mass energy of the collider is 500 GeV
with a luminosity of up to about $2.5\times 10^{-34}/cm^2/s$.   
In this energy region,
the production of a light Higgs boson is expected, in addition to the
high statics productions of Top quarks and $W$ bosons.  
The productions of other new particles are also predicted 
in models such as SUSY models.
Studies of these particles at JLC will make an indispensable contribution 
to our understanding of the fundamental forces and constituents.

Considering the importance of the $e^+e^-$ linear collider project, 
the Asian Committee for future Accelerators (ACFA) has initiated a
working group to study physics scenarios and experimental feasibilities at the linear collider 
in 1997\cite{ACFAHEP}.  Four ACFA workshops have been held since then,
and the group 
published a report in summer 2001\cite{ACFAReport}.
This report consists of discussions on physics at JLC, 
studies of a detector for JLC, and optional experiments using
$\gamma-\gamma$, $\gamma-e^-$, and $e^-e^-$ collisions.
Since it is impossible to cover everything in a short time, 
selected topics from the report are presented in this talk.

In the following two sections, the parameters of the JLC accelerator 
and the detector 
are described. In the subsequent section, selected topics 
concerning the physics on Higgs, Top and SUSY
are described. 
 Please see the ACFA report\cite{ACFAReport}, for topics not covered here.

\section{Accelerator}

\begin{figure}
\centerline{
\resizebox{0.8\textwidth}{!}{\includegraphics{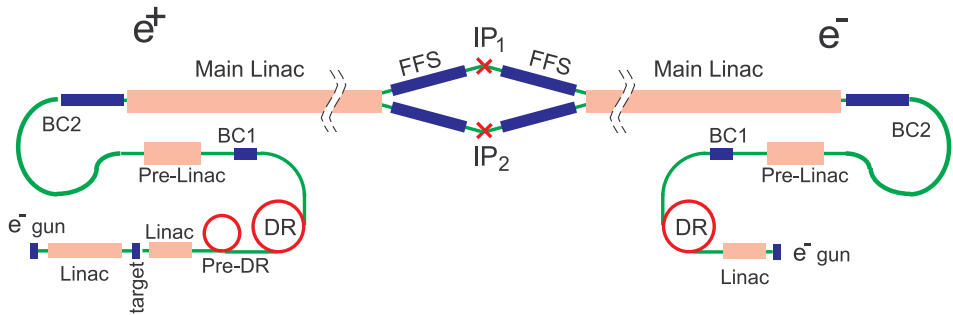}}}
\caption{\label{jlclayout.eps}{\it JLC Accelerator layout.}}
\end{figure}

The layout of the JLC accelerator is shown in Fig.~\ref{jlclayout.eps}.
The accelerator consists of two systems: one for electrons and the 
other for positrons.
In the electron system, electrons are generated by an $e^-$ {\it gun} and
accelerated to about 2 GeV by the {\it Linac}; 
the emittance is reduced by the damping ring ({\it DR}) and the longitudinal
beam size is reduced by the bunch compressors ({\it BC1, BC2}).
The electrons are then 
accelerated to by the {\it Pre-Linac} (up to 10 GeV),  and then by the
{\it Main Linac}. The final focus system ({\it FFS}) focuses
the beam size at the interaction points ({\it IP}).
The positron system is similar to the electron one, except for positron
generation.  Positrons are generated by injecting high energy 
electrons on a high Z target.  For efficient collection of the 
produced positrons,
an additional dumping ring ({\it Pre-DR}) is inserted before {\it DR}.

The parameters of the JLC accelerator are summarized in 
Table~\ref{jlcparameter}.  The first three columns (A, X, Y)
are parameters described in the ACFA Report\cite{ACFAReport}.
The A is a typical parameter set which we can expect at the early
stage of accelerator operation.
Together with the operation, tuning of the machine is proceed, 
and we will be able to operate high current beams with very low 
emittances, and the luminosity as high as the Y parameter can be
expected.    Recently, the international Technical Review Committee has
been organized to review accelerator technology developed in each region.
The last two columns in Table~\ref{jlcparameter}
show the parameters prepared for that committee.  
The TRC-500 parameter is similar 
to the Y parameter as for as the luminosity is concerned.

\begin{table}
\begin{center}\begin{minipage}{0.8\textwidth}
\caption{
\label{jlcparameter}{\sl
Parameters of the JLC accelerator. 
$^{\dag)}$ {\it ACFA Report 2001}, 
$^{\ddag)}$ {\it Technical Review Committee JLC/NLC(X)}\cite{Yokoya}.
{\it TRC(C) is in preparation }.}
}
\end{minipage}\end{center}
\begin{center}
\newfont{\tbfont}{ptmr8r}
{ \tbfont 
\begin{tabular}{|l | c | c | c | c | c | c |}
\hline
 & & A$^{\dag)}$ & X$^{\dag)}$ & Y$^{\dag)}$ & 
\multicolumn{2}{c|}{TRC(X)$^{\ddag)}$} \\
\hline
Center-of-mass energy & GeV & 535 & 497 & 501 & 
500 & 1000 \\
\hline
Repetition rate & $Hz$ &
\multicolumn{4}{c|}{150} & 100\\
\hline
Luminosity & $10^{33}/cm^2s$ & 
9.84 & 15.48 & 27.0 & 25.0 & 25.0\\
Nominal luminosity & $10^{33}/cm^2s$ &
6.82 & 11.15 & 18.20 & 15.2 & 15.7 \\
Integrated luminosity/year($10^7s$) & $fb^{-1}$ &
{98} & {155} & {270} & {250}
& {250} \\ 
\hline
\hline
Bunch separation & $nsec$ &
2.8 & \multicolumn{4}{c|}{1.4} \\
\hline
Bunch charge & $10^{10}$ &
0.75 & 0.55 & 0.70 & \multicolumn{2}{c|}{0.75} \\
Loaded gradient & $MV/m$ &
59.7 & 54.2 & 50.2 & \multicolumn{2}{c|}{55} \\
\hline
No. of bunches/pulse & &
95 & \multicolumn{2}{c|}{190} & \multicolumn{2}{c|}{192} \\
\hline
Linac length/beam & $km$ &
\multicolumn{2}{c|}{5.06} & 5.50 & 6.3 & 12.9 \\
\hline
AC power (2 linacs) & $MW$ &
\multicolumn{2}{c|}{118} & 128 & 150 & 200 \\
\hline
\hline
Bunch length & $\mu m$ &
\multicolumn{2}{c|}{90} & 80 & \multicolumn{2}{c|}{110} \\
\hline
Emittance at IP ($\gamma\epsilon_x / \gamma\epsilon_y$) & 
$10^{-8}m$-$rad$ &
400/6.0 & \multicolumn{2}{c|}{400/4.0} & \multicolumn{2}{c|}{360/40} \\
Beta function at IP($\beta_x/\beta_y$) & $mm$ &
10/0.10 & \multicolumn{2}{c|}{7/0.08} & 8/0.11 & 13/0.11 \\
\hline
Beam size at IP($\sigma_x/\sigma_y$) & $\mu m$ &
277/3.39 & 239/2.57 & 239/2.55 & 243/3.0 & 219/2.3 \\
Beamstrahlung energy loss ($\delta_B$) & \% &
4.42 & 3.49 & 5.22 & 4.7 & 8.9  \\
No. of photons per $e^-/e^+$ & &
1.10 & 0.941 & 1.19 & 1.3 & 1.3 \\
\hline
\end{tabular}
}
\end{center}
\end{table}

From an experimental point of view,
the JLC accelerator has several unique features.
Firstly, the JLC beam has a special time structure.
One pulse of the JLC beam consists of many bunches 
with a small time separation of 2.8 nsec to 1.4 nsec.
The pulse length of the beam is about 270 nsec, 
while there is a  very long interval on the order of 10 msec between 
each pulse.  This makes it hard to make an event trigger during a 
beam pulse while leaving sufficient time for a trigger decision 
between the pulses.

\begin{figure}
\centerline{
\resizebox{!}{0.25\textheight}{\includegraphics{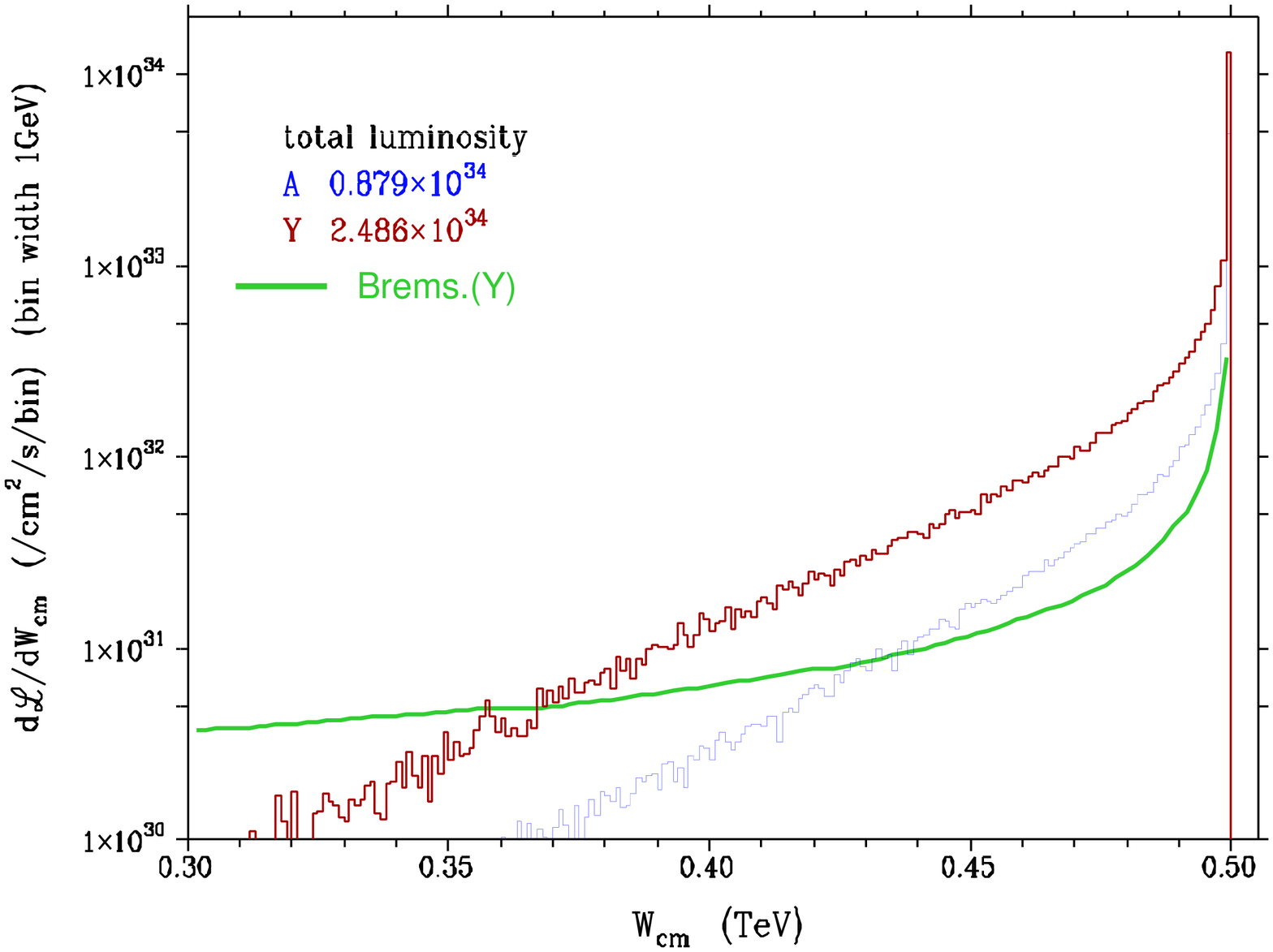}}
\hspace{0.05\textwidth}
\resizebox{!}{0.25\textheight}{\includegraphics{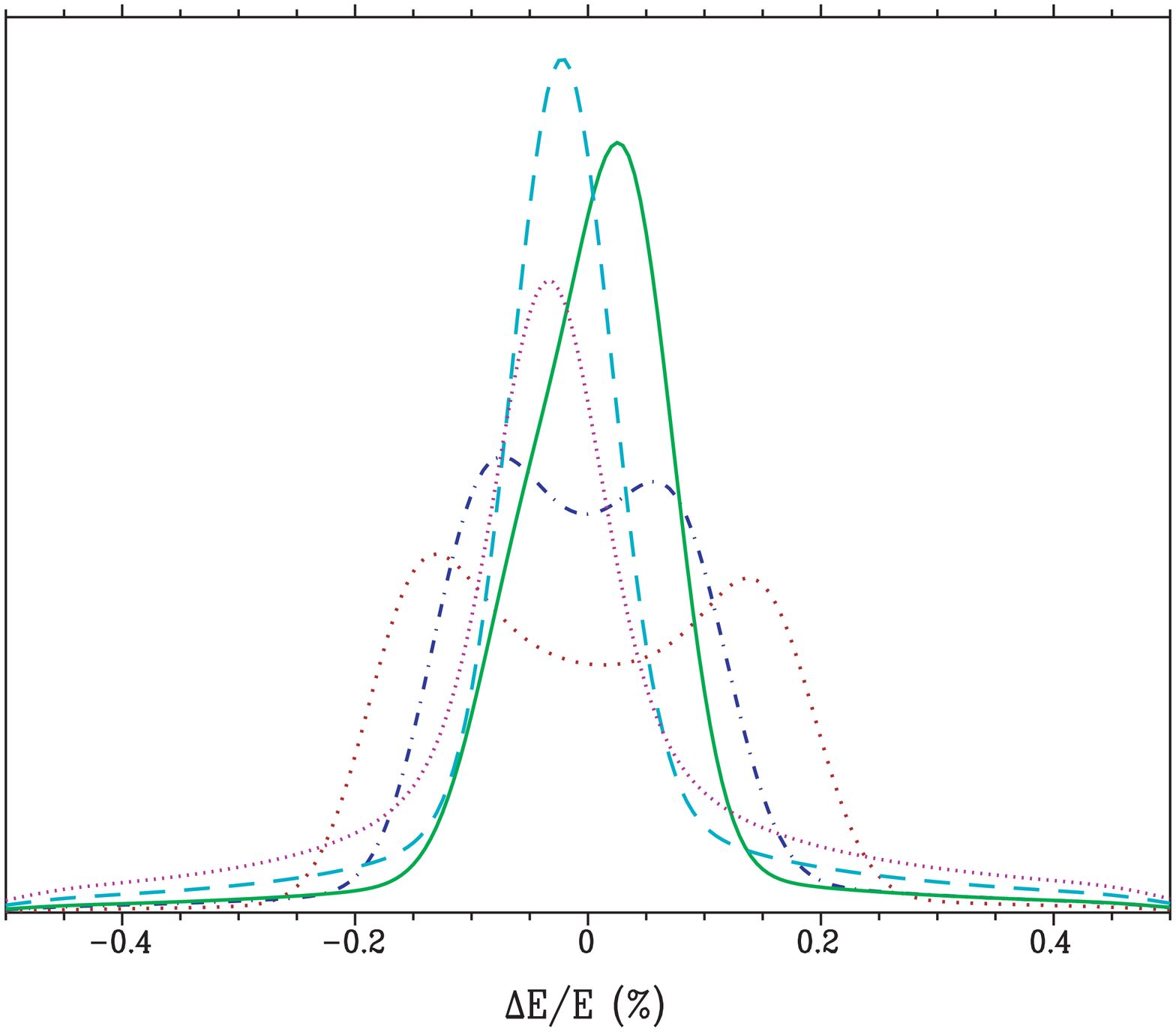}}
}
\begin{center}
\begin{minipage}[t]{0.40\textwidth}
\caption{\label{lum.eps}{\it Bremsstrahlung and beamstrahlung spectrum}.}
\end{minipage}
\hspace{0.12\textwidth}
\begin{minipage}[t]{0.40\textwidth}
\caption{\label{EnergySpread.eps}{\it 
Example of the beam energy spread at 250~GeV 
for various RF phase combinations.}}
\end{minipage}
\end{center}
\end{figure}

Secondly, due to the magnetic field produced by the 
opposite beam of high charge density, 
beam particles loose their energy and emit low energy 
$e^\pm$ and $\gamma$ particles, which is known as 
{\it Beamstrahlung}.  The beamstrahlung smears the 
collision energies and reduces the effective luminosity
at the nominal center-of-mass energy.  
The amount of this effect depends on the 
accelerator parameters, and it becomes significant just above 
the threshold of particle production where 
the production cross section changes rapidly with the energy.
The spectra in the case of A and Y parameters are 
shown in Fig.~\ref{lum.eps} 
together with the bremsstrahlung spectra.

\begin{figure}
\centerline{
\resizebox{0.7\textwidth}{!}{\includegraphics{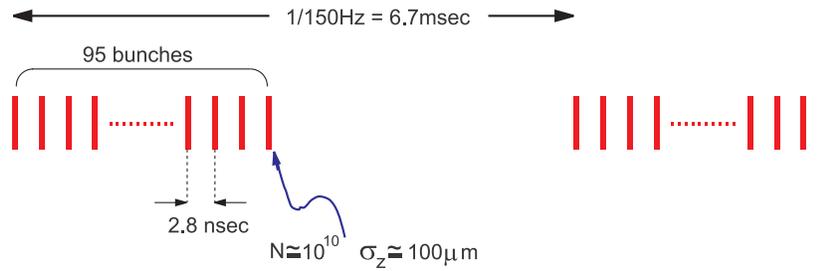}}}
\caption{\label{beampulse.eps}{\it Structure of the JLC beam pulse.}}
\end{figure}

Thirdly, the energy spread of the beam in the main linac 
is relatively larger than that of conventional circular colliders,
since there are no strong bending magnets to fix the energy.
However, the spread of the collision energy can be controlled to some 
extent by adjusting the RF phases  of the linacs.  We will be able to 
select the spectrum such as a wide but short tail spectrum, or a narrow  
but long tail spectrum, depending on the physics requirements.
The typical spectrum is shown in Fig.~\ref{EnergySpread.eps}.

Lastly, if bypass lines along the main linac are prepared,  
we can take data at lower energies without significant 
changes to the facility.
A run at the $Z_0$ pole is useful for detector calibration.
The cross sections  of $W$ pair production and top quark production 
are maximum at their threshold.  The same is true for the production 
of the light Higgs 
boson.  Since the energies for them are different, 
possibility to conduct experiments at different energies is important 
to maximize the physics output of the JLC.

\section{Detector}

The goals of the JLC detector performance have been set as follows:

\begin{itemize}
\item
Efficient, high-purity $b/c$ tagging capability.
\item
A momentum resolution of the tracking devices sufficiently good so that
the missing mass measured in the process
$e^+e^-\rightarrow ZH \rightarrow \ell\bar{\ell}X$ is only 
limited by the spread of the initial beam energy.
\item
The resolution of the calorimeter is good enough 
so that the invariant mass of a 2 jets system is comparable 
with that of the natural width of $W$ and $Z$.
\item
Good hermeticity so that invisible particles 
can be identified efficiently.
\item
Beam related backgrounds must be shielded by the masking system 
completely.  
The detector must be equipped with a time-stamping 
device so as to identify the bunch of the event and to reduce the
backgrounds which come from different bunches, but in the same beam pulse.
\end{itemize}

\begin{figure}
\centerline{
\resizebox{0.7\textwidth}{!}{\includegraphics{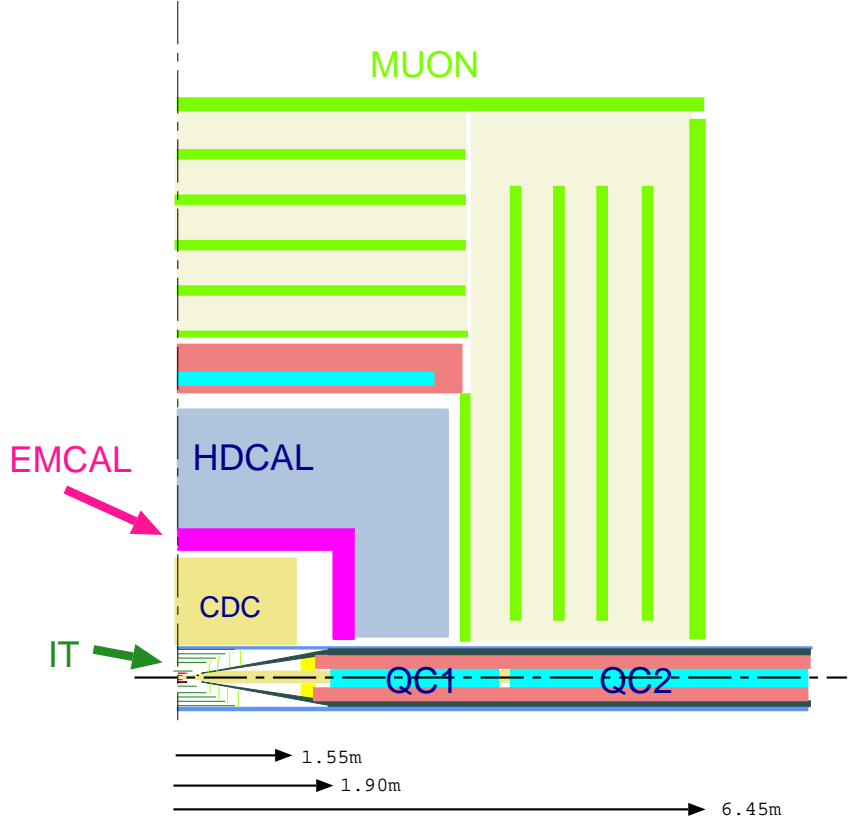}}}
\begin{center}
\begin{minipage}[t]{0.7\textwidth}
\caption{\label{det-xsect.eps}{\sl 
Cross section of the JLC detector.}
}
\end{minipage}
\end{center}
\end{figure}

\begin{figure}
\centerline{
\resizebox{0.7\textwidth}{!}{\includegraphics{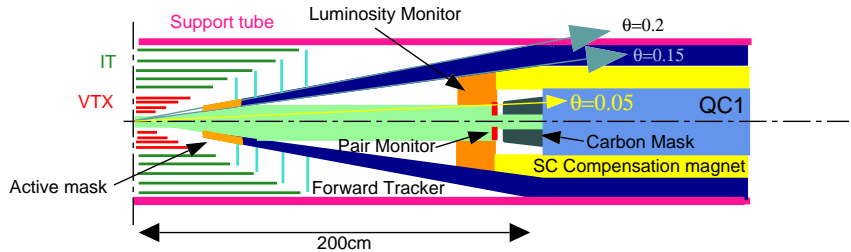}}}
\begin{center}
\begin{minipage}[t]{0.8\textwidth}
\caption{\label{det-ip.eps}{\sl 
The JLC detector system near IP}
}
\end{minipage}
\end{center}
\end{figure}

To this end, a general purpose detector was considered as a basis 
for physics studies and the development of detector technologies.
A cross-sectional view of the detector is shown in Fig.~\ref{det-xsect.eps},
and the detector system near the interaction 
region is shown in Fig.~\ref{det-ip.eps}.  
All of the detectors except for the muon detector are placed 
inside a solenoidal
magnetic field of 3 Tesla.  With a sandwiched lead-scintillator system,
the calorimeter is aimed to achieve a 
resolution of $15\%/\sqrt{E}({\rm GeV})\oplus 1\%$ 
for electro-magnetic particles and $40\%\sqrt{E}({\rm GeV})\oplus 2\%$ for 
hadronic particles.   A central tracking device is 
a small cell jet chamber.  Together with a CCD vertex detector, 
a momentum resolution($\frac{\sigma_{p_t}}{p_t}$) of 
$1\times 10^{-4} p_t ({\rm GeV})\oplus 0.1\%$ is expected.
In the forward region, the region above 200 mrad is fully covered by 
the calorimeter.  For the region below 200 mrad, an active mask, 
a luminosity monitor and a pair monitor will be used to measure 
energetic $e^+/e^-$; thus, the minimum veto angle is 11 mrad.
Please see the ACFA report\cite{ACFAReport} concerning 
detector hardware studies performed to achieve these design goals.

\section{Physics}

\begin{figure}
\begin{center}
\resizebox{!}{0.5\textheight}{\includegraphics{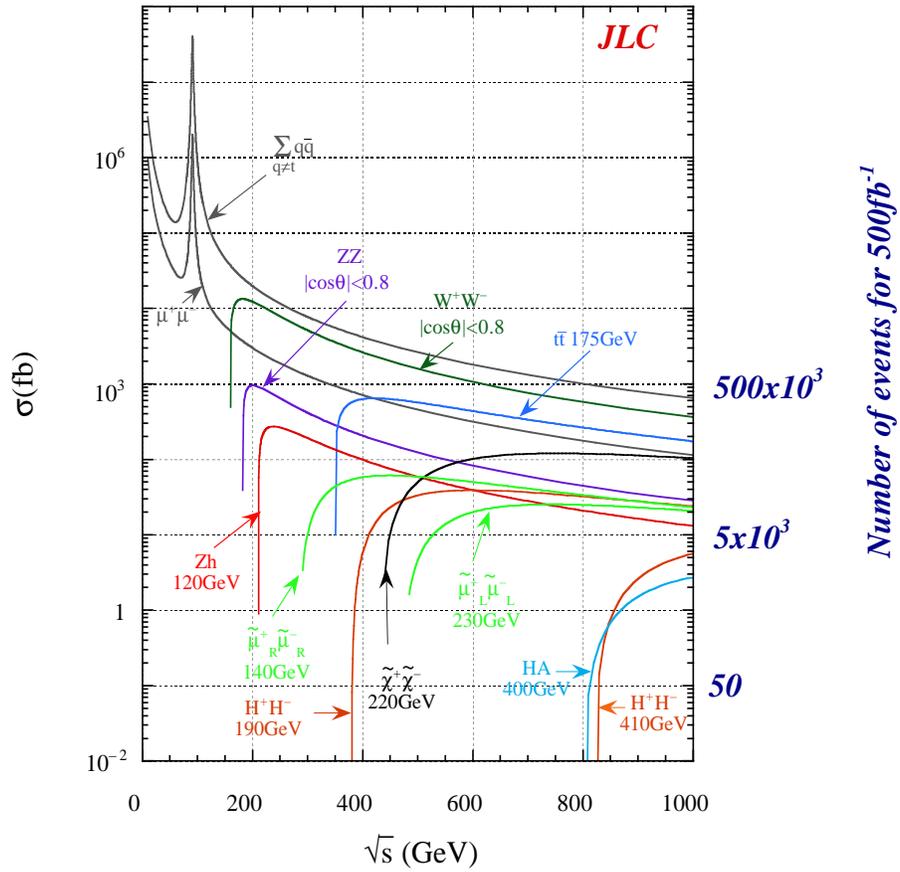}}
\end{center}
\begin{center}
\begin{minipage}[t]{0.8\textwidth}
\caption{\label{crosssections.eps}{\it 
Cross sections of standard model processes 
and new particle productions. The number of events corresponding to 
an integrated luminosity of 500 $fb^{-1}$ is indicated on the right.}}
\end{minipage}
\end{center}
\end{figure}

The total cross sections of the standard model processes and 
new particle productions 
are shown in Fig.~\ref{crosssections.eps}.  With the machine parameters of 
Y in Table~\ref{jlcparameter}, we can expect to collect data of 
the integrated luminosity of 500 fb$^{-1}$ within two years. 
This will allow us not only discoveries of 
new particles but also precise studies 
of new particles and the standard model particles.  
The expected precisions that we can obtain from these studies 
and the implications for physics 
beyond the standard model are described in the ACFA report in detail.
Here, selected topics on Higgs, Top and SUSY physics are
described in the following subsections.

\subsection{Higgs}
\label{higgs-section}

The standard model of elementary particles consists of three families of 
matter fermions, gauge bosons 
and Higgs boson. Gauge bosons are massless due to gauge symmetry, but
they acquire masses when Higgs spontaneously breaks the symmetry.
Though the gauge nature of forces among fermions has been tested very
precisely experimentally\cite{GAUGETEST}, 
the key feature of the standard model, 
that the symmetry is broken 
by the Higgs, has yet to be confirmed experimentally.
To this end, first of all, the Higgs particle must be discovered.

In the standard model, the mass of the Higgs boson is just a parameter.
However, its self energy diverges with its mass.
It is so divergent that if its mass is heavier than about 200 GeV, 
new physics must show up below the GUT energy and GUT models 
must be abondened.  If it is light, 
the Higgs can be elementary up to the GUT energy.

\begin{figure}
\begin{center}
\resizebox{0.7\textwidth}{!}{\includegraphics{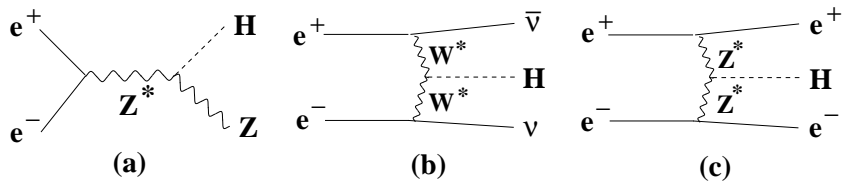}}
\end{center}
\begin{center}
\begin{minipage}[t]{0.8\textwidth}
\caption{\label{higgs-diagrams.eps}{\it 
Feynman diagram of Higgs production at JLC;
(a) Higgs-strahlung, (b) $WW$ fusion, (c) $ZZ$ fusion}}
\end{minipage}
\end{center}
\end{figure}

Although the Higgs boson has been searched extensively, 
no signal of direct production has been found,
and a lower mass bound of 114.1 GeV at the 95\% 
confidence level has been set\cite{HiggsMassBound}.
On the other hand, the precise measurements of the standard model processes 
allow us to probe tiny loop effects of the Higgs boson.
According to a global analysis of the electro-weak data, 
the most probable value of the Higgs boson mass is 
$106^{+57}_{-38}$ GeV while it is less than 222 GeV at 
the 95\% CL\cite{GAUGETEST}.
Thus the Higgs boson is likely to be light and its production is expected 
at an early stage of the JLC experiments.

\begin{figure}
\begin{center}
\resizebox{0.5\textwidth}{!}{\includegraphics{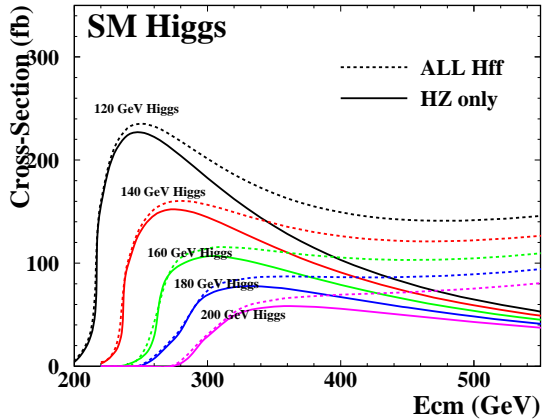}}
\end{center}
\begin{center}
\begin{minipage}[t]{0.8\textwidth}
\caption{\label{higgs-xsec.eps}{\it 
Total cross section of  Higgs production at the JLC energy region, 
for Higgs masses of 120, 140, 160, 180, and 200 GeV.
The cross sections of the Higgs-strahlung process are indicated 
by the solid lines, while those of all diagrams are 
shown by the dotted lines.}}
\end{minipage}
\end{center}
\end{figure}

Feynman diagrams of the Higgs production at JLC are shown in 
Fig.~\ref{higgs-diagrams.eps}.  They 
consists of Higgs-strahlung from $s$-channel $Z$ boson,
and $t$-channel productions by $WW$ and $ZZ$ fusion.
The total cross sections  of the Higgs 
production near to the threshold region are shown in Fig.~\ref{higgs-xsec.eps}.
For a Higgs of mass 120 GeV, the cross section is maximum at
a center-of-mass energy of about 250 GeV. 
For this case, its production is mainly by the Higgs-strahlung 
process,  while
at 500 GeV, about half of the cross section is due to the $WW$ and $ZZ$
fusion processes.

\begin{figure}
\begin{center}
\resizebox{0.3\textwidth}{!}{\includegraphics{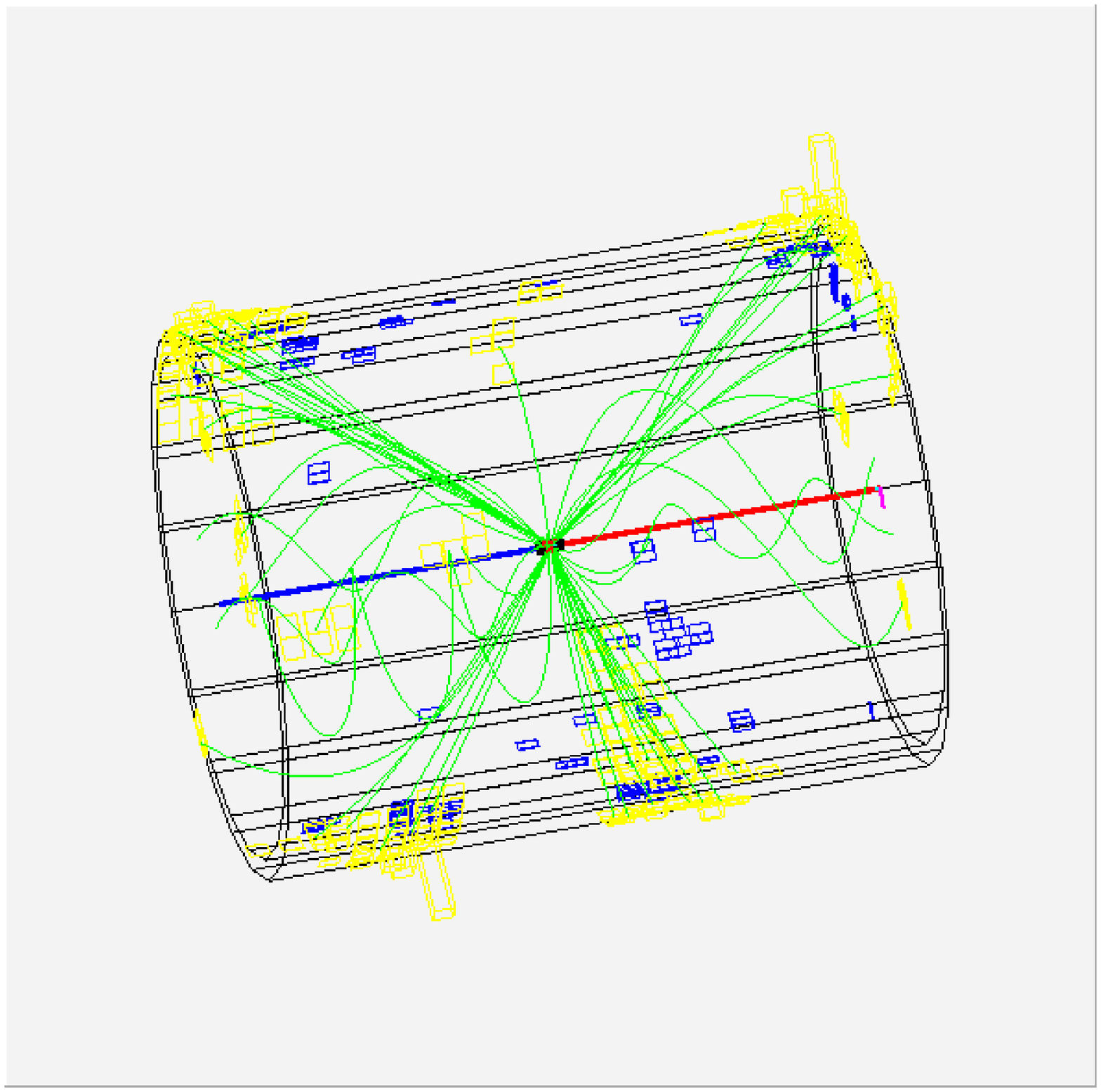}}
\resizebox{0.3\textwidth}{!}{\includegraphics{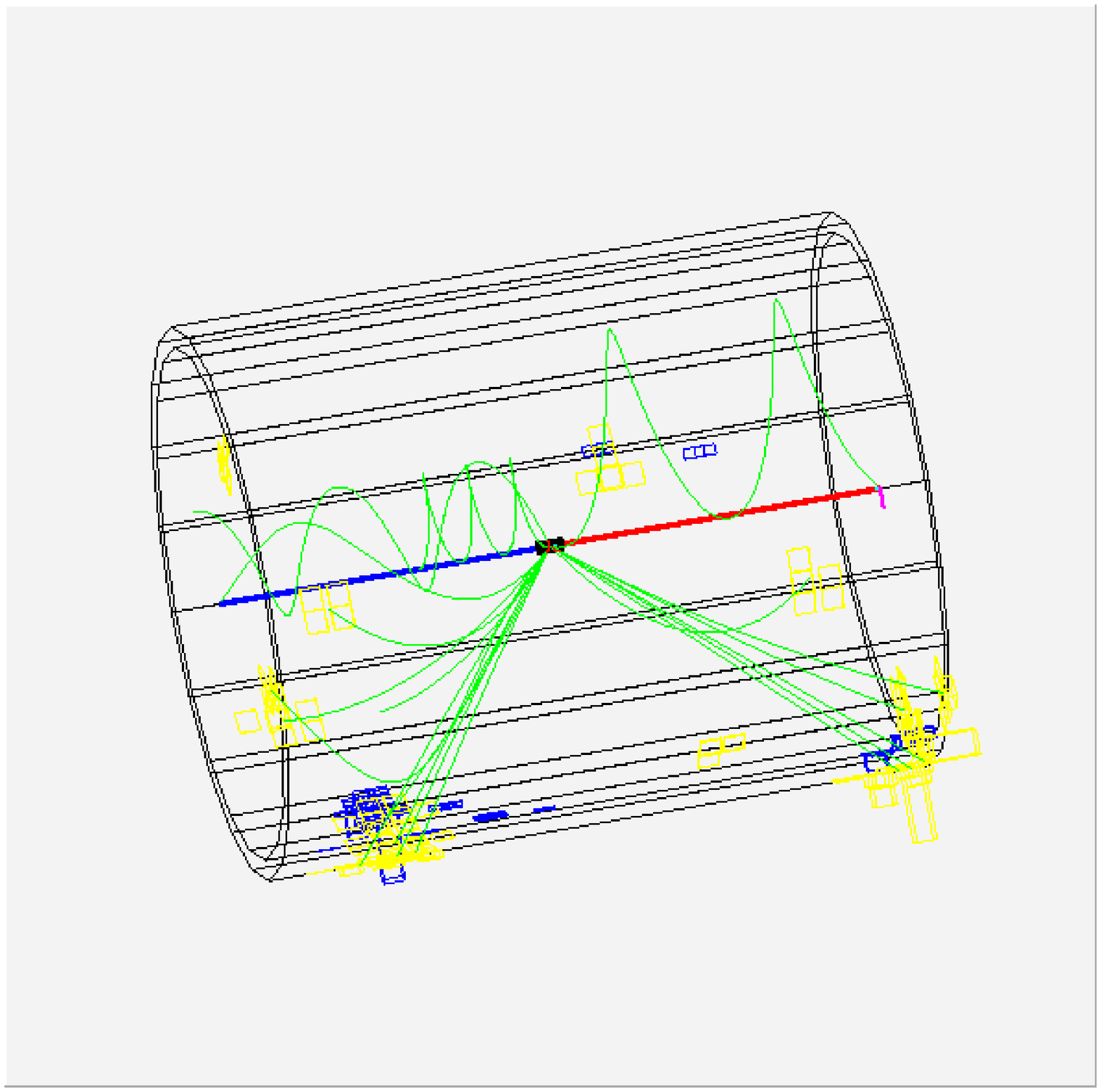}}
\resizebox{0.3\textwidth}{!}{\includegraphics{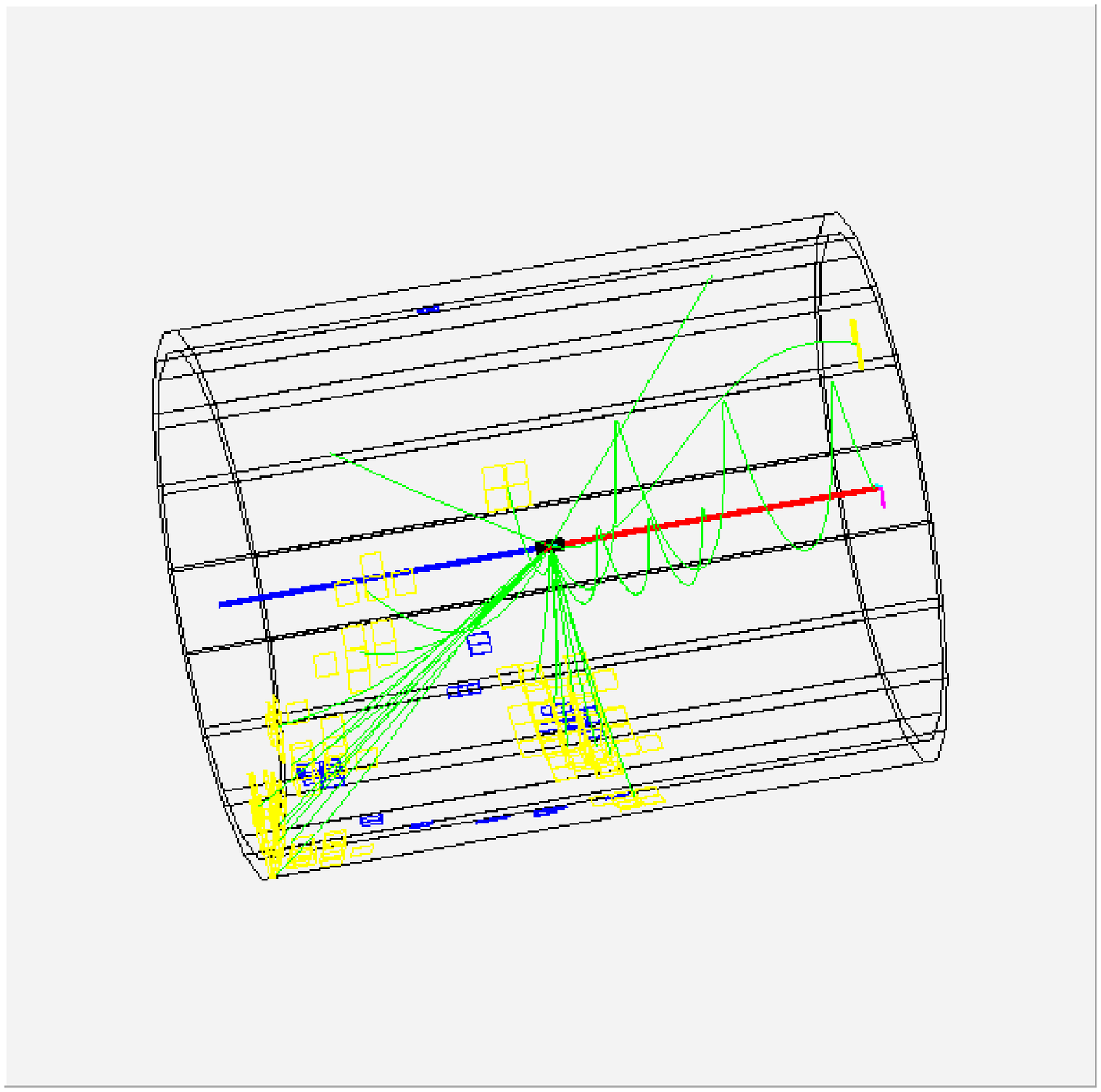}}
\end{center}
\begin{center}
\begin{minipage}[t]{0.8\textwidth}
\caption{\label{higgsevents}{\it 
Typical Higgs events; 4 jets, 2 jets + 
missing, and 2 jets + 2 leptons.
}}
\end{minipage}
\end{center}
\end{figure}

As an example of the Higgs study at JLC, 
we consider an experiment at $\sqrt{s}$=250 GeV
and a Higgs mass of 120 GeV.  For this case, the Higgs 
is produced mainly by Higgs-strahlung and
it decays mainly 
to $b\bar{b}$ (Branching ratio of $H\rightarrow b\bar{b}$ decay
is about 67\%).
Therefore the event signature of Higgs production is categorized, 
according to the 
decay mode of $Z$, to 4 jets, 2 jets + missing, or 2 jets + 2 leptons
as shown in Fig.~\ref{higgsevents}.  

\begin{figure}
\begin{center}
\resizebox{0.8\textwidth}{!}{\includegraphics{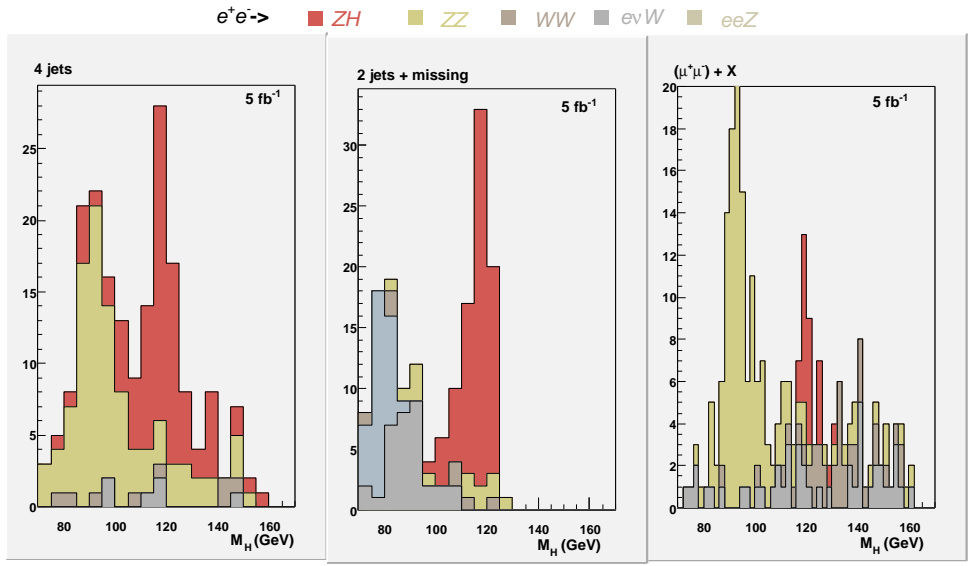}}
\end{center}
\begin{center}
\begin{minipage}[t]{0.8\textwidth}
\caption{\label{higgs-5fb}{\it 
Distribution of Higgs mass for an integrated luminosity 
of 5 fb$^{-1}$ measured in the 4-jet (left), 
the 2-jet + missing (center), and $\mu\bar{\mu}$ + missing (right) modes.
The Higgs process is shown in brown, and the backgrounds are shown 
by other colors.
}}
\end{minipage}
\end{center}
\end{figure}

The 4-jet mode is  selected by requiring, in principle, 
a successful forced four-jet clustering
and $b$-jet tagging. Selecting two jet-pairs, where the mass of one pair 
and the missing mass of the other pair are both consistent with $Z$, 
we could see a clear indication of Higgs even at a 
low integrated luminosity of 5 fb$^{-1}$
(Fig.~\ref{higgs-5fb}a).  In the case of the 2-jet mode, 
the mass of the Higgs is measured 
by just calculating the invariant mass of all of the detected particles.  
Requiring $b$-jet tagging, the missing $p_t$ and 
the missing mass being consistent with $Z\rightarrow \nu\bar{\nu}$, a clean
signal can also be seen in this mode (Fig.~\ref{higgs-5fb}b) at 
a integrated luminosity of 5~fb$^{-1}$. In Fig.~\ref{higgs-5fb}c, 
the missing mass of $\mu\bar{\mu}$ disregarding the decay mode
of the Higgs is shown. The selection criteria for this mode is just two $\mu$ tracks in good tracking device acceptance,
and that their invariant mass is consistent with $m_z$.
This mode allows a search independently of Higgs decay modes.  
If the Higgs mass is heavier than about 140 GeV, the Higgs decay to 
gauge bosons becomes dominant and the search in 4-jet and 2-jet modes must be replaced 
by searches in gauge boson modes.  However, a search in the 2-lepton mode is valid
even in this case.  Once the design luminosity of the JLC is achieved, 
the standard Higgs boson will be discovered very easily, as long as it is kinematically accessible.

\begin{figure}
\begin{center}
\resizebox{0.8\textwidth}{!}{\includegraphics{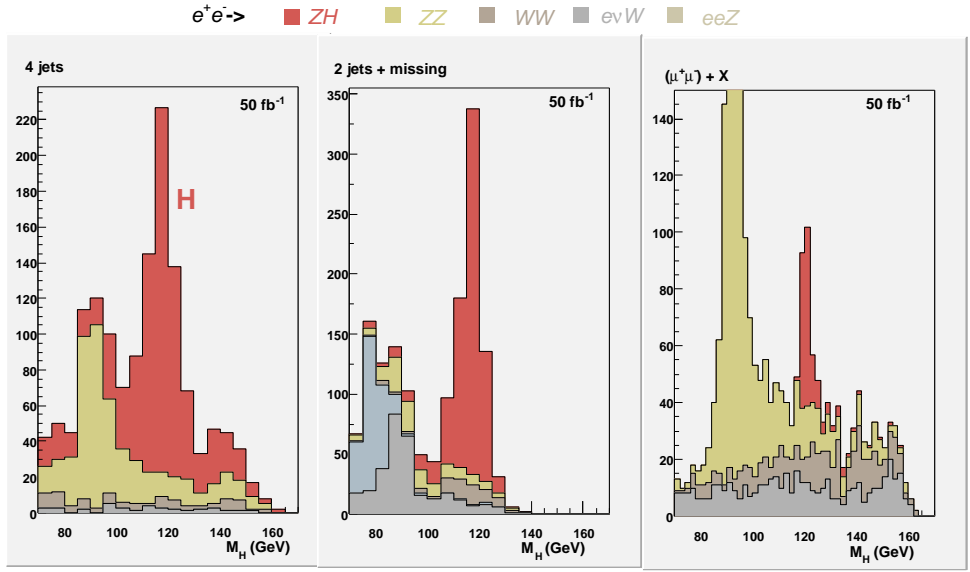}}
\end{center}
\begin{center}
\begin{minipage}[t]{0.8\textwidth}
\caption{\label{higgs-50fb}{\it 
Same as Fig.~\protect\ref{higgs-5fb}, but for the integrated luminosity 
of 50 fb$^{-1}$.
}}
\end{minipage}
\end{center}
\end{figure}

If 50 fb$^{-1}$ of data is taken, the Higgs signals are 
evident in all three decay modes,
as shown in Fig.~\ref{higgs-50fb}.
Note that 50 fb$^{-1}$ of data can be collected for less than 
one month of the design luminosity of parameter Y.   

\begin{figure}
\begin{center}
\resizebox{0.4\textwidth}{!}{\includegraphics{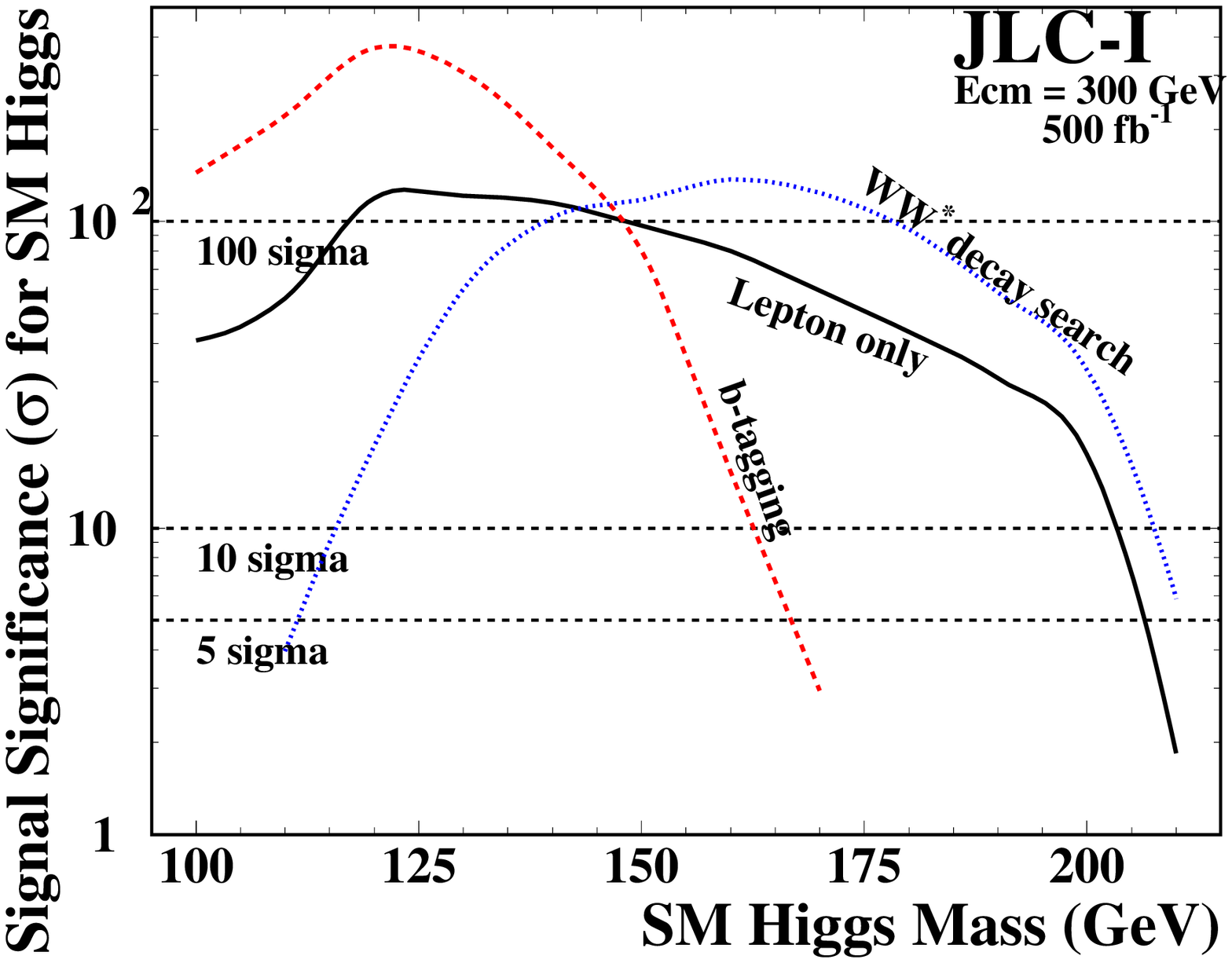}}
\resizebox{0.4\textwidth}{!}{\includegraphics{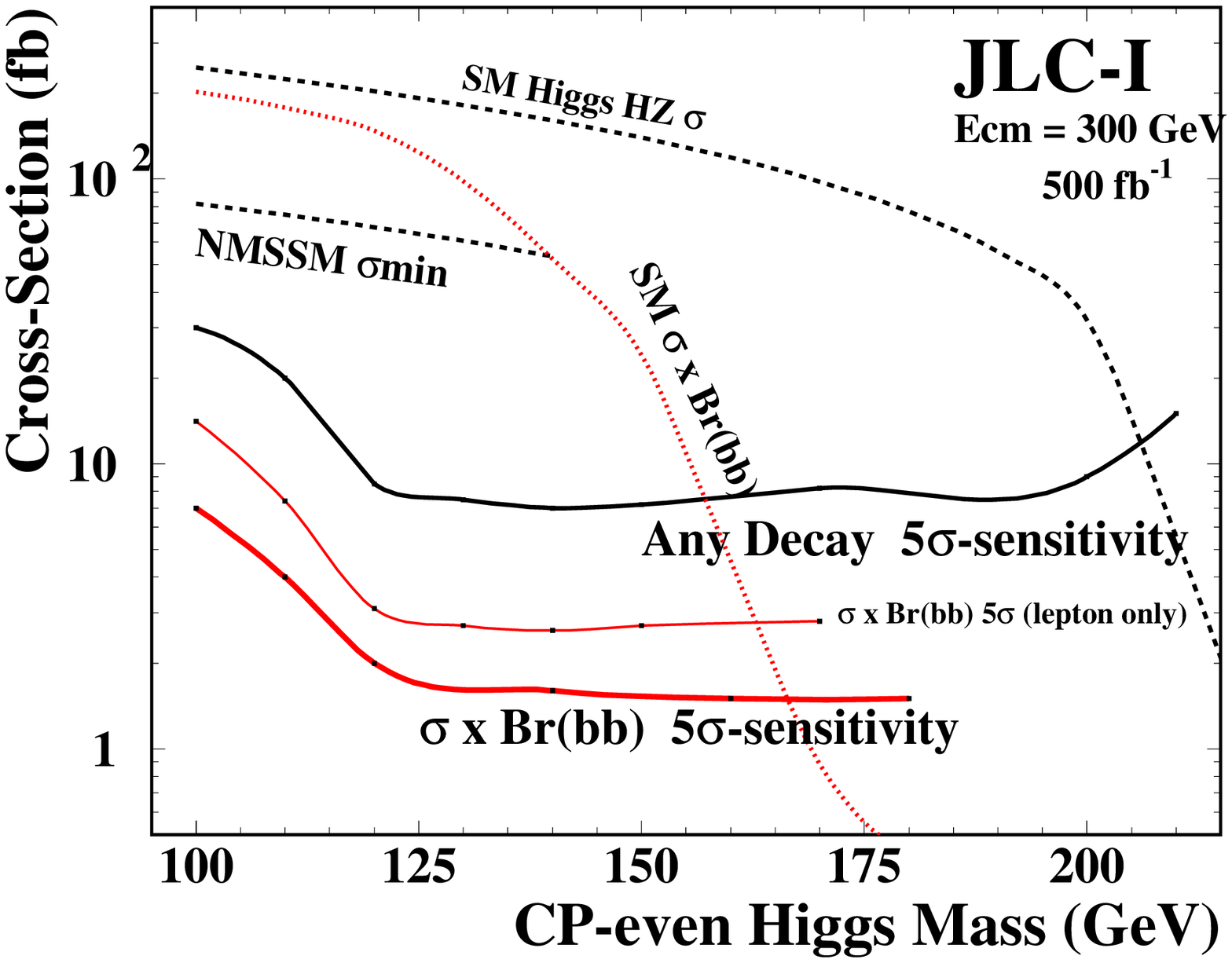}}
\end{center}
\begin{center}
\begin{minipage}[t]{0.8\textwidth}
\caption{\label{higgs-discovery}{\it 
Significances of the standard model Higgs when 500 fb$^{-1}$ is accumulated
(left) and cross sections of 5$\sigma$ significance
for non-standard CP-event Higgs boson (right).
}}
\end{minipage}
\end{center}
\end{figure}

With two years of operation of the Y parameter, we will be able to accumulate 
500~fb$^{-1}$ data.  With this amount of data, the significance 
for the standard model Higgs boson is more than 100 if its mass is light
(Fig.~\ref{higgs-discovery}).  Even if the standard model Higgs 
is not the case, a CP-even Higgs boson can  be searched 
irrespective to their decay mode, and if not found, 
the 5$\sigma$ upper bound of the cross section of 10 fb can be  obtained
for Higgs mass of up to 200 GeV.  This will rule out any GUT model
where the Higgs self coupling should not diverge up to GUT energy
.

\vskip 12pt
Once a new particle is discovered, the next task is to study 
its property, such as spin, parity and 
the strength of $ZZH$ coupling, and to establish the particle as the Higgs.
To study the spin, the threshold 
behavior\cite{HiggsSpin} and the angular distribution 
of the production and decay are useful.

To measure the mass of Higgs, three methods can be considered: (1) direct mass reconstruction 
using $H\rightarrow 2 jets$ mode, (2) a measurement of the recoil mass using the 
$Z\rightarrow e\bar{e}/\mu\bar{\mu}$
decay mode, and (3) combined fitting with beam-energy constraints.
The natural width of the Higgs is very small 
(about several MeV unless gauge boson channels dominate)
and JLC has a wide energy spread.   The detector must have a high resolution 
at least comparable with the beam energy spread, and be well calibrated using 
$Z$ signals for example.

\begin{figure}
\centerline{
\resizebox{!}{0.25\textheight}{\includegraphics{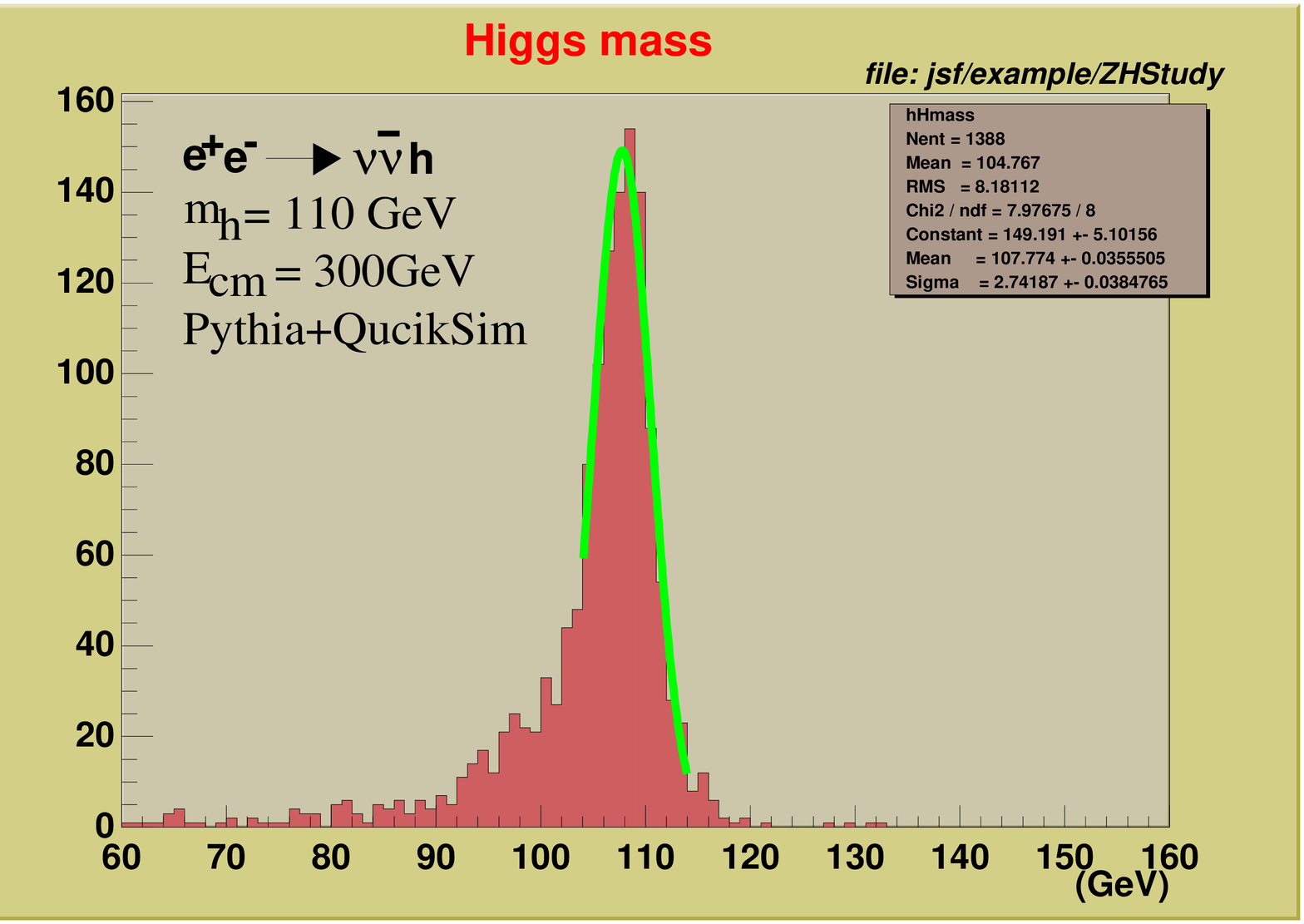}}
\resizebox{!}{0.25\textheight}{\includegraphics{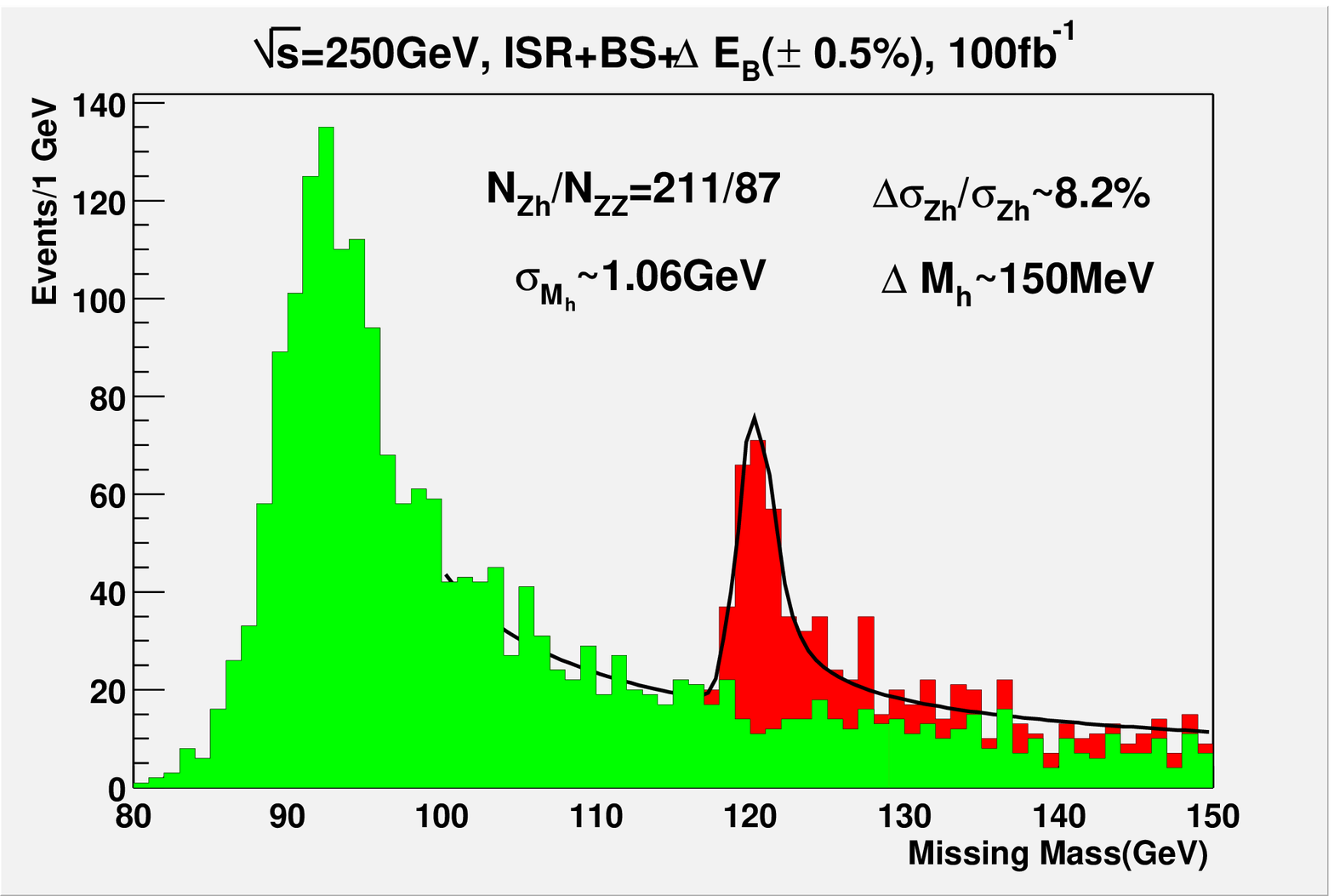}}}
\begin{center}
\begin{minipage}[t]{0.40\textwidth}
\caption{\label{hmass.eps}{\it
Mass distribution of the Higgs boson detected in 2-jet mode.
The invariant mass of all particles of selected events are plotted.
}}
\end{minipage}
\hspace{0.121\textwidth}
\begin{minipage}[t]{0.40\textwidth}
\caption{\label{MM.zhzz250.eps}{\it 
Recoil mass distribution of a $\mu^+\mu^-$ pair of
processes $e^+e^-\rightarrow ZZ$ and $e^+e^-\rightarrow ZH$,
at $\sqrt{s}=250$ GeV.  See the texts for more details.
}}
\end{minipage}
\end{center}
\end{figure}

A typical mass resolution obtained using the 2-jet mode 
is shown in Fig.~\ref{hmass.eps}.  This plot is obtained from an example program 
of the JSF package\cite{JSF}.  According to a Gauss fit, we obtained 
a $\sigma$ of 2.7 GeV, while the peak position is shifted about 2.3 GeV.
About 15k events are expected for 500 fb$^{-1}$.  Therefore, 
a statistical accuracy of less than 30 MeV for the Higgs mass is expected,
while the detector must be well calibrated for an unbiased measurement.

If we use the lepton channel, the mass measurement will be less biased 
because the Higgs is detected independently of its decay mode as the recoil 
mass of the lepton pair.  However, to achieve high precision measurements 
of the Higgs mass, a precise tracking device and 
a narrow initial energy spread is essential.
An example, in the case of the distribution of the recoil mass of 
$\mu$ pair is shown in Fig.~\ref{MM.zhzz250.eps}.
In this figure, the signal channel ($e^+e^-\rightarrow ZH$ for $m_H=120$ GeV)
is shown together with the background channel ($e^+e^- \rightarrow ZZ$).
Backgrounds below the signal peak are $ZZ$ events at lower collision energies
whose energy losses are due to not only the initial state radiation but also 
the beamstrahlung.  The backgrounds can be reduced siginificantly if we 
apply $b$-tagging to the jets from the Higgs decay, but it is not applied here 
for the measurement irrespective to the Higgs decay mode.
The width of the signal is mainly due to the initial 
beam energy spread, but is reduces by about 10\% if measured at a 10 GeV
lower center-of-mass energy due to a kinematical effect\cite{LCWS2000-TRACKING}.
In any case, a mass resolution of about 150 MeV and a precision of 
the cross section of about 8\%  are expected from this channel.
If we combine the $e^+e^-$ channel and the integrated luminosity of 500 fb$^{-1}$ 
is considered,  a mass resolution of about 45 MeV and 
a resolution of the cross section of about 2.7\%  are expected.

\vspace{12pt}
The branching ratio of the Higgs boson also needs to be measured, because
they may provide hints for physics beyond the standard model.
In the standard model, the strength of Higgs fermion coupling is 
proportional to the fermion mass.  However, many models beyond 
the standard model assumes more than one Higgs doublets; thus 
the proportionality of the mass and the coupling strength obeys different formula.

When the Higgs decays to a fermion pair, 
the vertex detector is very powerful to identify its flavor.
Several methods, such as $n-sig$ and topological vertex hunting 
algorithms have been studied and found to be useful to identify 
Higgs to $c\bar{c}$ decay mode\cite{gbyu} at a precision of 
25 \%.

The Higgs decay to $WW^*$ was studied in the process  
$e^+e^-\rightarrow ZH$, where $Z$ decays to $q\bar{q}$ or $\ell^+\ell^-$
and $H$ decays to $WW^*\rightarrow \ell\nu q\bar{q}$. For this study, 
good jet mass measurements are crucial and lepton channels are useful to reduce 
combinatorial backgrounds. With the help of a kinematical constraint fit,
5.1\% accuracy of $\Delta(\sigma\cdot Br(H\rightarrow WW))/\sigma\cdot Br(H\rightarrow WW^*)$
is expected for an integrated luminosity of 500~fb$^{-1}$.

In the standard model, the total width of the Higgs boson is 2 to 10 MeV 
for a light Higgs boson of mass 
from 100 to 140 GeV, thus a directly measurement of the width is difficult.
However, by combining the measurement of $\sigma Br(H\rightarrow WW)$ with the measurement of 
the total cross section $(\sigma_{total})$, 
the total width $(\Gamma_{total})$ of the Higgs boson can be estimated.
This is because  $\Gamma_{total}=\Gamma(H\rightarrow WW)Br(H\rightarrow WW)$ and 
$\Gamma(H\rightarrow WW)$ can be obtained from $\sigma_{total}$ assuming 
$\Gamma_{HZZ}/\Gamma_{HWW} = (M^W\cos\theta_W / M_Z)^2$, which is satisfied 
as long as $W$ and $Z$ are SU(2) gauge bosons.  The accuracy of the 
measurement of the total width can be expressed as
$$
\left( {\Delta\Gamma_H \over \Gamma_H} \right)^2 = 
\left( {\Delta \sigma\cdot BR(H\rightarrow WW)} \over \sigma\cdot BR(H\rightarrow WW)\right)^2 
+ \left( 2{\Delta \sigma_{total}\over \sigma_{total} } \right)^2,
$$
which becomes about 6.4\% for an integrated luminosity of 500~fb$^{-1}$.
Since heavy particles make a non-negligible contribution
to the total width, this measurement provides a tool to prove 
the particle spectrum at higher energies.

\subsection{Top}

According to the PDG\cite{PDG}, the mass of the top quark is 
$174.3\pm 3.2\pm 4.0$ GeV.  This implies that the JLC 
at $\sqrt{s}\sim $ 350 GeV would be a top factory.
In addition, because $m_t$ is larger than $m_W+m_b$, 
the top quark decays to $Wb$ almost 100\% of the time in the standard model.
The total width ($\Gamma_t$) is about 1.4 GeV, which implies that 
the top decays before entering the non-perturbative QCD region.
$\Gamma_t$ acts as an infrared cutoff, and clean tests of QCD are possible.
Vertecies including top, such as 
$t\bar{t}\gamma$, $t\bar{t}Z$, $tbW$, $t\bar{t}g$, and $t\bar{t}H$,
could be studied, where effects due to new physics beyond the standard
model may be found.

At $t\bar{t}$ threshold, $t\bar{t}$ resonances are formed. 
The lowest resonance ($E_{1S}$) is about 3 GeV below  twice the 
mass of the top quark. There are many resonances below the open threshold,
while their mass differences are small, less than, say, 2-times the total width 
of the top quark.  Therefore, although they overlap, a ``bump'' in the cross section 
will be seen. From its position in $\sqrt{s}$, we will know the mass of the top quark.
From its height and width, information such as the total width of the top quark, 
and $\alpha_s$ can be obtained.  
On the other hand, the initial state radiation and the beamstrahlung reduce 
the luminosity usable for resonance production, and the initial beam energy spread 
smears the peak and/or affects the shape of the peak if the relative spread is 
larger than about 0.4\%.

When a top quark is produced, it decays to $bW$. $W$ decays to a quark pair 
or a lepton and a neutrino.  According to the decay mode of $W$, the signature 
of $t\bar{t}$ events are categorized as follows:
\begin{center}
\begin{tabular}{l l l}
  &  & Branching ratio \\
a) & 2 $b$ jets + 4 jets from $W$ & $\sim 45$\%, \\
b) & 2 $b$ jets + 2 jets from $W$ and $\ell\nu$ & $\sim 44$\%, \\
c) & 2 $b$ jets + 2 pairs of $\ell\nu$ & $\sim 11$\%.
\end{tabular}
\end{center}
All modes can be used for measuring the total cross section.
The direction and charge of the top quark can be identified in 
decay modes a) and b), which are used to measure the top quark momentum 
and the forward-backward asymmetry.
The basic cuts to select $t\bar{t}$ events are:
(1) event shape cuts such as those on the number of charged particles,
the number of jets and thrust, (2) mass cuts to select $W$'s and b jets,
(3) requirements of leptons in the cases of b) and c), and (4) $b$ tagging.

\begin{figure}
\begin{center}
\resizebox{\textwidth}{!}{\includegraphics{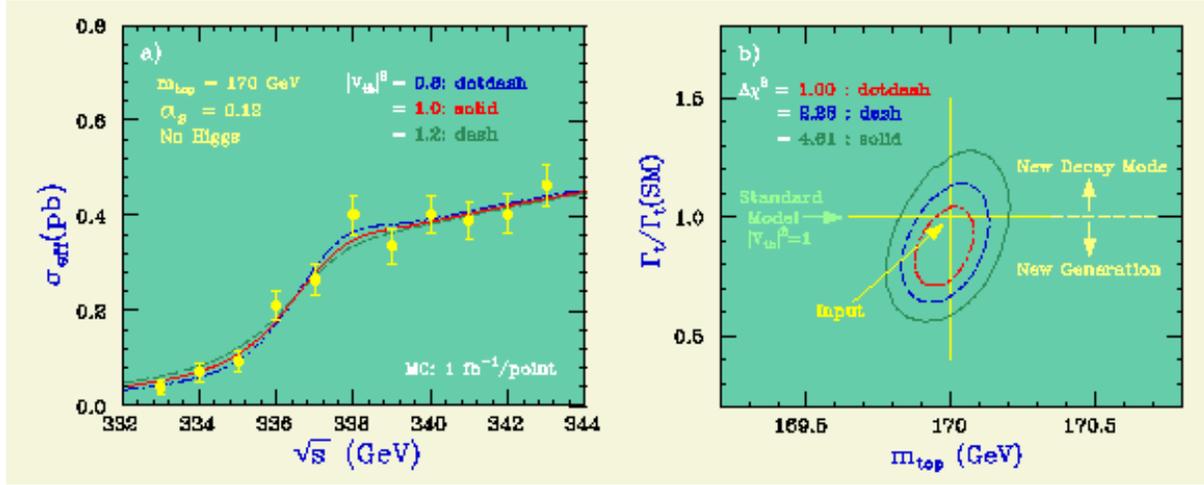}}
\end{center}
\begin{center}
\begin{minipage}[t]{0.8\textwidth}
\caption{\label{mtvtb2.eps}{\it 
Energy dependence of the $t\bar{t}$ cross section(a) 
and the expected precision in the plane of the top mass and the 
total width(b). The 11 point measurements around the threshold, 
with an integrated luminosity of 1~fb$^{-1}$ each are assumed.
}}
\end{minipage}
\end{center}
\end{figure}

A typical measurement of the energy dependence of the $t\bar{t}$ cross section around 
the threshold region is shown in Fig.~\ref{mtvtb2.eps}-(a). In the figure, 11 points of 1~fb$^{-1}$ 
measurement are shown.  The lines in the figure are theoretical curves for 
three cases for the value of $|V_{tb}|$. If $|V_{tb}|$ becomes smaller, $\Gamma_t$ becomes smaller 
and the threshold becomes steeper.  The position of the shoulder is determined by the mass of the top.
Thus, this measurements could impose a constraint on the plane of the total width and the mass 
of the top quark, as shown in Fig.~\ref{mtvtb2.eps}.

\begin{figure}
\begin{center}
\resizebox{0.8\textwidth}{!}{\includegraphics{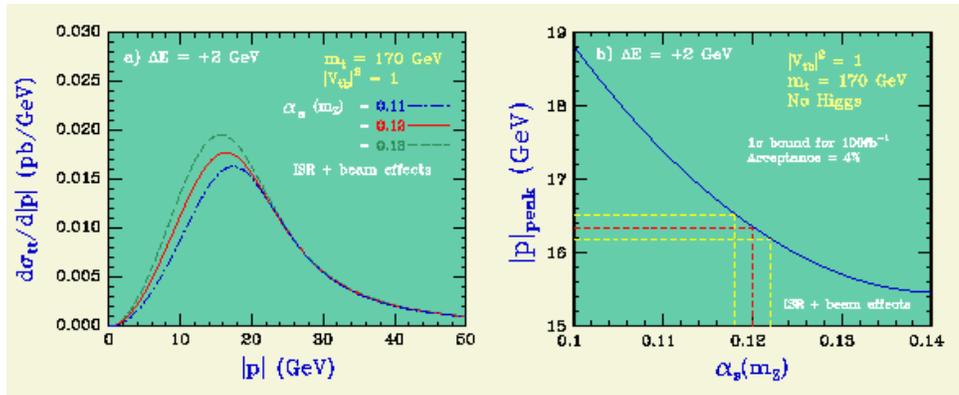}}
\end{center}
\begin{center}
\begin{minipage}[t]{0.8\textwidth}
\caption{\label{ppkalfs.eps}{\it 
Momentum distribution of the top quark for various values of $\alpha_S$ (a) 
and the peak momentum as a function of $\alpha_s$ (b).
}}
\end{minipage}
\end{center}
\end{figure}

When the toponium resonance is formed, the top quark decays to $bW$ before 
$t\bar{t}$ annihilation.  Therefore, if the $t\rightarrow bW$ decay is measured 
precisely, the top quark momentum in the resonance can be reconstructed to study
the toponium potential.  This is in contrast to the charmonium and the bottomnium resonance 
where the $q\bar{q}$ annihilation modes dominate.  In Fig.~\ref{ppkalfs.eps}, the momentum distribution of 
the top quark in the toponium resonance is shown for three values of $\alpha_s$.
The larger is $\alpha_s$, the deeper is the toponium potential and the larger is the 
peak momentum.  According to a simulation study, 1$\sigma$ bound for the precision of 
the peak momentum $(|p|_{peak}) $ measurement  is about 200 MeV for an 
integrated luminosity of
100~fb$^{-1}$.  This value translates to sensitivities of $\Delta\Gamma_t / \Gamma_t=0.03$
and $\Delta\alpha_s(M_Z)/\alpha_s=0.002$, when only one 
parameter is varied while the others is fixed 
and the mass of the lowest toponium resonance is known.

\begin{figure}
\begin{center}
\resizebox{!}{0.3\textheight}{\includegraphics{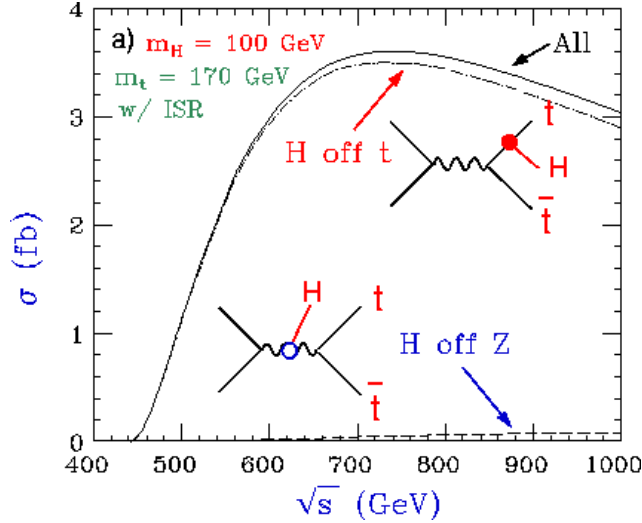}}
\end{center}
\begin{center}
\begin{minipage}[t]{0.8\textwidth}
\caption{\label{sgtth.eps}{\it 
Total cross section of the process $e^+e^-\rightarrow t\bar{t}H$.
}}
\end{minipage}
\end{center}
\end{figure}

\vspace{12pt}

At energies higher than the threshold of the process $e^+e^-\rightarrow t\bar{t}H$, 
we can measure the top Yukawa coupling directly.  There are two kinds of diagrams 
contributing to this process at the tree level.  The first one includes a diagram where a Higgs boson
is radiated off the final-state top or anti-top quarks $(H-off-t)$, and is thus proportional to the top
Yukawa coupling.  In the second diagram, the Higgs boson is emitted from the $s$-channel Z boson
$(H-off-Z)$ and is independent of the top Yukawa coupling. However, the contribution of 
the $H-off-Z$ diagram is tiny, as shown in Fig.~\ref{sgtth.eps}.

Considering the light Higgs, where it decays mainly to $b\bar{b}$, 
the signature of this process is 2 $W$'s and 4 $b$'s, 
where $W$ decays to $q\bar{q}'$ or $\ell\nu$.
In a simulation study, 8 jets or 6 jets + $\ell\nu$ final states
were selected using jet clustering to find $W$ and $H$ jet pairs.
The background processes are $e^+e^-\rightarrow t\bar{t}$ 
and $e^+e^-\rightarrow t\bar{t}Z$ .  
The former process can be removed 
by requiring  more  than two $b$-jets. 
The latter process is irreducible, and the background is severe 
if the Higgs mass is close to the mass of $Z$.
If $m_H=100$ GeV and $m_t$=170 GeV, the expected number of events 
at $\sqrt{s}=700$ GeV is 114, combining 8-jet and 6-jet + $\ell\nu$ mode, while 
that of the background process is 133, when the integrated luminosity is 100 fb$^{-1}$.
This translates to an accuracy of the top Yukawa couping of 14\%.

\subsection{SUSY}

Since the radiative correction to the scalar particle is quadratically divergent, 
the light Higgs boson in GUT models raises the fine-tuning problem\cite{finetune}.
One motivation of SUSY is to solve this problem by introducing super-symmetric particles 
with mass on the order of 1 TeV or less, and cancel the divergence.
Another solution to the fine-tuning problem is to introduce extra dimensions\cite{ExtraDimension}.

Obviously, SUSY is broken and the mass spectrum of SUSY 
particles varies depending on the SUSY breaking models and model parameters\cite{ACFAReport}.
Since SUSY particle searches at JLC is model independent, their measurements
are useful to distinguish various models.

In gravity mediated models,
gaugino masses are unified at the GUT scale, but at the EW scale,
charginos and neutralinos are lighter than the gluino ($m_{\tilde{\chi}} \sim \frac{1}{3} m_{\tilde{g}}$)
because they receive radiative corrections differently.
Accordingly, the lightest neutralino ($\tilde{\chi}_1^0$) is expected to be the lightest SUSY particle.
Similarly, the right-handed fermion is expected to be the lightest matter fermion
($m_{\tilde{f}_R} < m_{\tilde{f}_R} < m_{\tilde{q}}$).  
Therefore, channels to search for sparticles are the  right-handed sfermion($\tilde{f}_R$) decays 
to the right-handed fermion($f_R$) and LSP($\tilde{\chi}^0_1$) or the chargino decays
($\tilde{\chi}^\pm$) to 
the gauge bosons($W^\pm$) and LSP($\tilde{\chi}^0_1$).
The event signature  of the sparticle production 
is a missing $p_t$.

In anomaly mediated models, SUSY symmetry is broken by loop effects.
The mass differences among gauginos and fermions are usually small, though
the neutralino is expected to be LSP.  If the mass difference of LSP and the next LSP
is large, the signature of the sparticle production is a missing $p_t$. It is similar to the 
case of the gravity mediated models.  However, if the mass difference is small,
particle multiplicities  of events are small and special care must be taken to 
find sparticles.

On the contrary, in the gauge mediated models, the gravitino is the LSP and 
the next lightest sparticles, NLSP, can be 
$\tilde{\chi}_1^0$, lightest stau ($\tilde{\tau_1}$), or right-handed electron ($\tilde{e}_R$) 
depending on model parameters.
Their lifetimes also depend on the parameters.
If the lifetime of the NLSP is short and decays near to the interaction point, 
the  event signature is the missing $p_t$ due to energies taken away by the gravitino.
If it is long, but decayed,  with in the detector,
spectacular events, such as off-vertex $\gamma$ or $\tau$ tracks may be seen.
If their lifetime is very long and  does not decay within the detector,
the NLSP momentum is not detected.  In this case, we needs to search  for the next-to-next 
LSP to NLSP decays,  and the event signature is again the missing $p_t$.

\begin{figure}
\centerline{
\resizebox{0.7\textwidth}{!}{\includegraphics{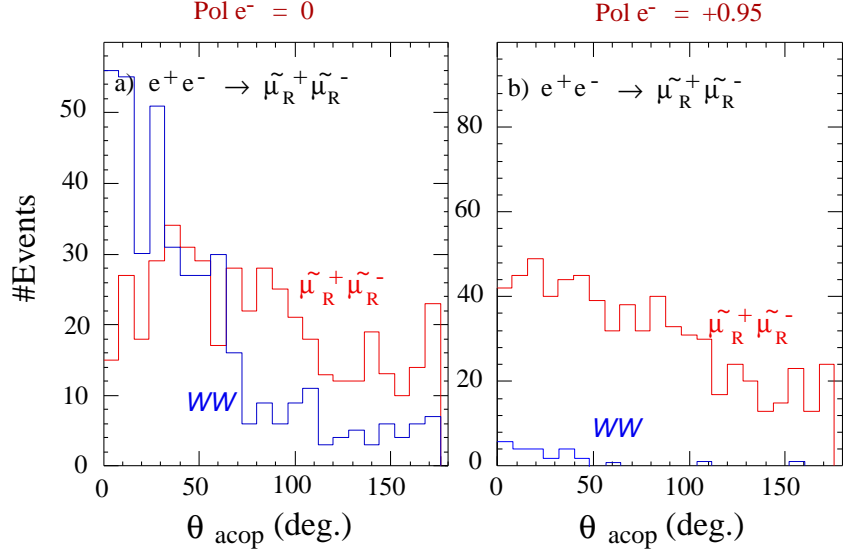}}
}
\begin{center}
\begin{minipage}[t]{0.8\textwidth}
\caption{\label{smusmu.eps}\it
Examples of the acoplanarity distribution of smuon pair production
for without a polarized beam (a) and with a polarized $e^-$ beam.
The Monte Calro data corresponds to an integrated 
luminosity of 20~fb$^{-1}$ at $\sqrt{s}=350$ GeV.
}
\end{minipage}\end{center}
\end{figure}

\begin{figure}
\centerline{
\resizebox{0.8\textwidth}{!}{\includegraphics{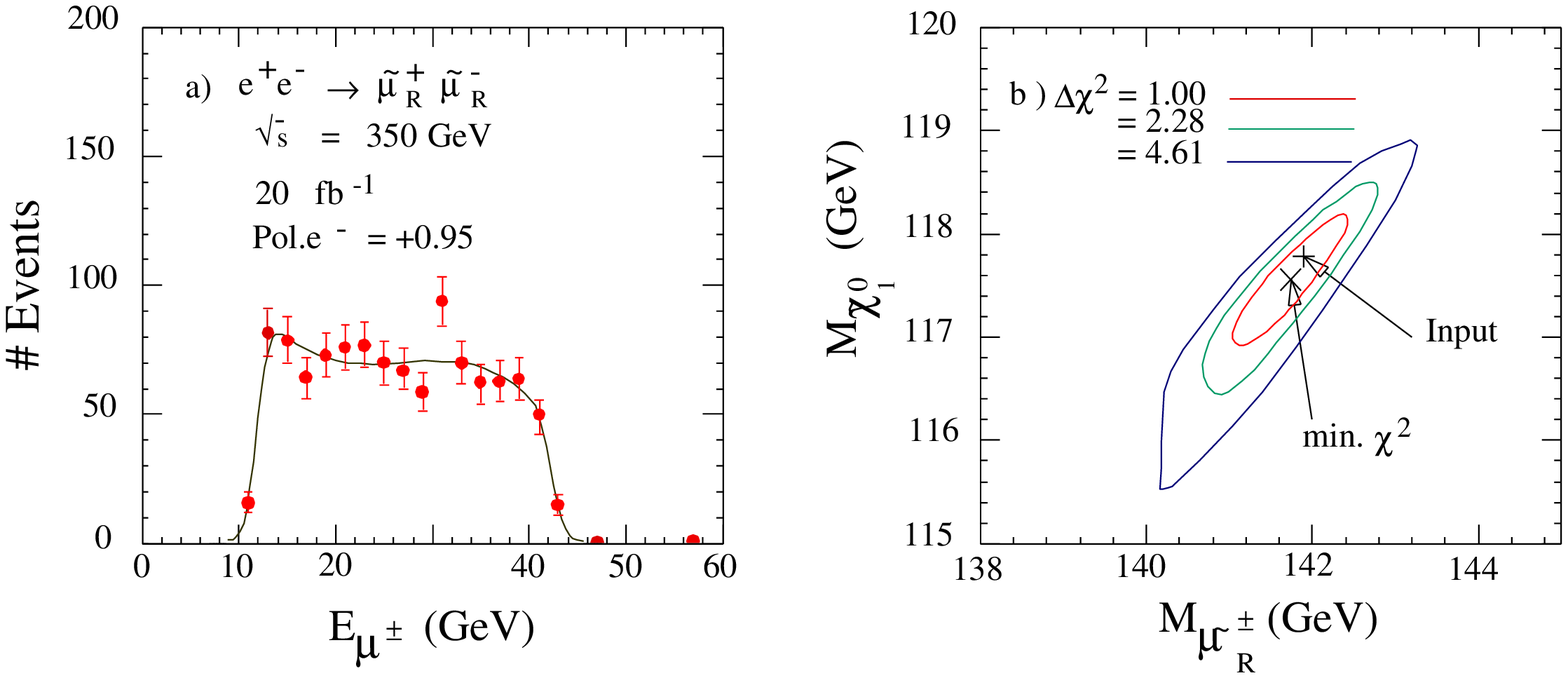}}
}
\begin{center}
\begin{minipage}[t]{0.8\textwidth}
\caption{\label{smu_mlmlsp.eps}{\it
(a) Energy distribution of muons from smuon decays 
and (b) contours in the $m_{\tilde{\mu}}-m_{\tilde{chi}^0_1}$
plane obtained from a fit to the energy distribution.
}}
\end{minipage}\end{center}
\end{figure}

In any case, the detection of SUSY particle is easy at JLC once the kinematical threshold
is exceeded, and no model assumption, such as mass spectrum etc,
is required for its detection. 
Once they are discovered, the masses and couplings will be precisely measured 
from which we will study the SUSY breaking mechanism and underlying physics.

As an example of SUSY studies at JLC, searches and studies of right handed scalar
leptons are discussed below.  We consider the 
process, $e^+e^-\rightarrow \tilde{\mu}^+ \tilde{\mu}^-$,
where $\tilde{\mu}^\pm$ decays to $\mu^\pm$ and $\tilde{\chi}^0_1$.
The signature of this event is a missing $p_t$ due to an un-detected momentum of $\tilde{\chi}^0_1$,
which leads to acoplanar $\mu^\pm$ events.
The distribution of the acoplanarity angle($\Theta_{acop}$) is shown in Fig.~\ref{smusmu.eps}.
Since only 
the $B$ boson (the gauge boson of $U(1)_Y$ group) contributes to smuon($\tilde{\mu}^\pm$) production,
the use of a right-handed electron beam enhances the signal twice, while it reduces
the major $W^+W^-$ backgrounds completely, as can be seen in the Figure.

In the decay of smuon, the energy distribution of muon is flat 
as smuon is a scalar particle. The end points of its energy distribution
are kinematically fixed by the masses of the parent and daughter particles.
Namely, the masses of smuon and the LSP can be determined from the end points 
of the smuon energy distribution.  An example of the energy distribution is shown 
in Fig.~\ref{smu_mlmlsp.eps}.  
For the case used for a Monte Calro study, 
the expected mass resolution
of smuon is $\pm 0.8$ GeV, and that for LSP is 
$\pm 0.6$ GeV if the 20~fb$^{-1}$ data is corrected at $\sqrt{s}=350$ GeV.

If the universal scalar mass is the case such as expected in models like
gravity mediated models, not only the smuon but also the selectron is produced at a similar energy.
In this case, the signature is acoplanar $e^\pm$,  and can be easily distinguished from an
acoplanar $\mu^\pm$ signal.  The mass of selectron is measured from the end points
of $e^\pm$ energy distribution.  Combining the mass measurement of smuon,
the model assumption of the universal scalar mass can be easily tested.

\begin{figure}
\centerline{
\resizebox{0.4\textwidth}{!}{\includegraphics{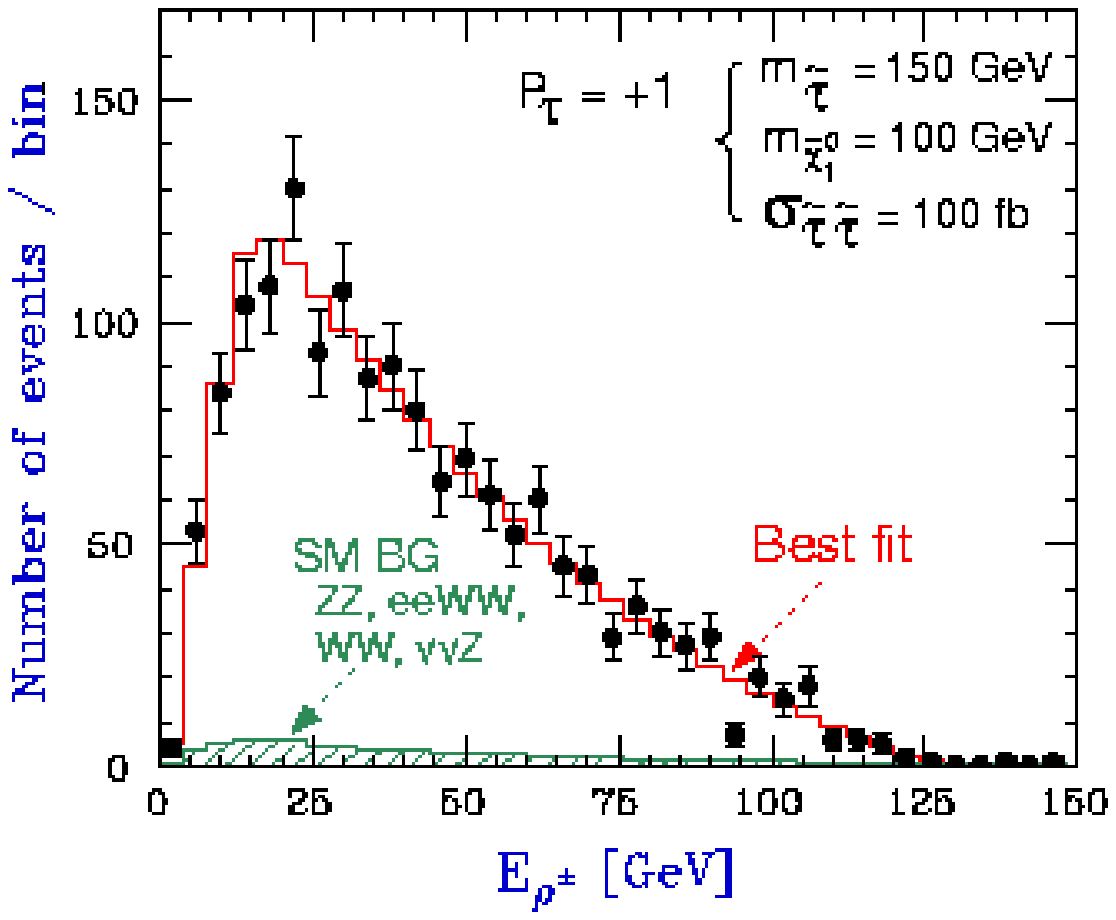}}
\hspace{0.05\textwidth}
\resizebox{0.4\textwidth}{!}{\includegraphics{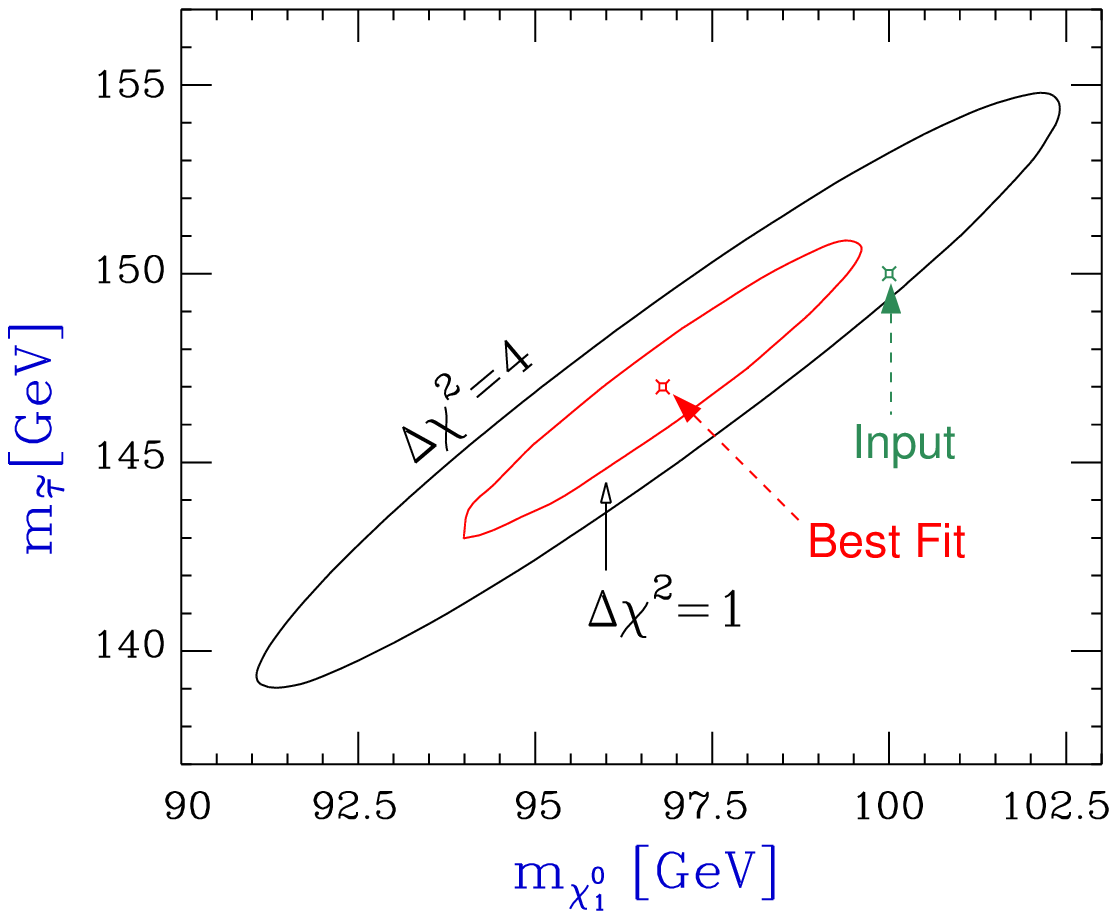}}
}
\begin{center}
\begin{minipage}[t]{0.80\textwidth}
\caption{\label{stau_massfit.eps}{\it
a) Energy distribution of the final state hadrons selected as $\rho$'s 
from stau decays shown together with the best fit, when 
$10^4$ $\tilde{\tau}_1$ pairs are produced and decay 
in the $\tilde{chi}^0_1 \tau_R$ mode,
and b) $\Delta\chi^2=1, 4$  contour in the $m_{\tilde{\chi}^0_1} - m_{\tilde{\tau}_1}$
mode.}}
\end{minipage}\end{center}
\end{figure}

Concerning the third generation slepton, stau($\tilde{\tau}$), 
its mass matrix contains large off-diagonal elements since the
tau mass ($m_\tau$) is much heavier than the other leptons.
This produces a large mixing between left-handed and right-handed 
stau and a large mass splitting between the mass eigen states of the stau ($\tilde{\tau}_1$,
$\tilde{\tau}_2$) is expected.  Thus,
$\tilde{\tau}_1$ may be the lightest charged SUSY particle.

The signature of stau pair production is a pair of leptons or low multiplicity 
hadron jets from tau decays.  Since a neutrino is produced in tau decay,
the energy distribution of the observed leptons or hadron jets is not uniform, 
as with the case of the smuon or selectron.  Still, the energy distribution of the
decayed daughter reflects the masses of the stau and neutralino and we will be able 
to determine their masses.  An example of an energy distribution of  $\rho$ mesons
from tau decay is shown in Fig.~\ref{stau_massfit.eps}-(a).  From the fit, 
the mass of the stau can be determined at 2\% precission using a 
sample corresponding to an integrated luminosity of 100 fb$^{-1}$.

Generally, the total cross section of the right handed slepton and the left handed 
slepton is different due to the difference of  the weak hyper charge.
If  $\sqrt{s}>\!\!>m_Z$ and the beam electron is right-handed,
only a $B$ boson of the $U(1)_Y$ gauge group contributes to slepton production.  In this case,
the total cross section of the right handed slepton is four times larger than 
that of the left handed slepton.  Thus if the mass is fixed, the cross section of the 
lightest stau($\tilde{\tau}_1$) is determined by the mixing ratio of 
$\tilde{\tau}_R$ and $\tilde{\tau}_L$ ($\sin\theta_{\tilde{\tau}}$).
To put it another way, the mixing angle of the stau sector can be measured 
using a measurement of the total cross section of a stau.
According to a monte calro simulation of 100 fb$^{-1}$ luminosity 
at 500 GeV, a 6.5\% precission measurement of $\sin\theta_{\tilde{\tau}}$ is 
expected.

There is also a mixing in the stau decay, which decays to a tau and a neutralino.
The neutralino consists of a
Bino($\tilde{B}^0$), a Wino($\tilde{W}_3^0$) and Higgsinos($\tilde{H}_1^0$ , $\tilde{H}_2^0$).
If a stau interacts with a Bino or a Wino, a tau with the opposite helicity is produced.
If a stau interacts with a Higgsino, a tau with the same helicity is produced.
As a result, the polarization of the daughter tau is determined 
by the mixing angles of the neutralino and the stau mixing angle. This property can be 
exploited to determine $\tan\beta$\cite{staupol}.

\section{Summary}

In this talk, after a short summary of the JLC accelerator and the detector, 
selected topics on physics of the Higgs, Top and SUSY particles are presented.
The initial goal of JLC is to achieve a center-of-mass energy of 500 GeV
with a  luminosity of about $2.5\times 10^{34}/cm^2/s$.   
JLC will be  a Top factory.  If the Higgs boson is as light as expected from 
recent electroweak data, JLC will also be a Higgs factory.  
Some SUSY particles may be observed at JLC.  
A clean experimental environment will allow us to provide definite results
on studies of these particles. It will be an indispensable basis for 
our understanding of the physics beyond the standard model.
The beginning of the JLC experiment is highly awaited.

\vspace{24pt}
\noindent
{\bf Acknowledgments}\\
This talk is based on the ACFA report\cite{ACFAReport}.
The author would like to thank the members of ACFA Joint Linear Collider
Physics and Detector working group.


\end{document}